\documentclass[a4paper, 11pt]{article}
\usepackage[a4paper, total={17cm, 24cm}]{geometry}

\usepackage{amsthm, amsmath, amsfonts}
\usepackage{caption}
\usepackage{physics}
\usepackage{tikz} %
\usetikzlibrary{external, arrows.meta, positioning, bending, backgrounds}
\usetikzlibrary{quantikz2}
\usepackage{pgfplots}
\pgfplotsset{compat=1.18}

\title{Quantum data encoding as a distinct abstraction layer in the design of quantum circuits}

\author{Gabriele Agliardi$^{*,1,2}$ and Enrico Prati$^{\dag,3,4}$\\[8mm]
{\footnotesize\parbox{.8\textwidth}{
$^*$ gabriele.agliardi@it.ibm.com $^\dag$ enrico.prati@unimi.it\\[2mm]
$^1$ IBM Quantum, IBM Research, Via Circonvallazione Idroscalo I--20090 Segrate, Milano, Italy\\
$^2$ At the time of paper conception, also at Dipartimento di Fisica, Politecnico di Milano, Piazza Leonardo da Vinci I--20133 Milano, Italy\\
$^3$ Dipartimento di Fisica “Aldo Pontremoli”, Università degli Studi di Milano, Via Celoria 16, I–20133 Milano, Italy\\
$^4$ Istituto di Fotonica e Nanotecnologie, Consiglio Nazionale delle Ricerche, Piazza Leonardo da Vinci 32, I–20133 Milano, Italy\\
}}
}

\theoremstyle{definition}
\newtheorem{defn}{Definition}[section]

\theoremstyle{plain}
\newtheorem{prop}[defn]{Proposition}

\theoremstyle{remark}
\newtheorem{remark}[defn]{Remark}

\DeclareMathOperator{\poly}{poly}

\begin{document}
\maketitle

\abstract{
Complex quantum circuits are constituted by combinations of quantum subroutines. The computation is possible as long as the quantum data encoding is consistent throughout the circuit. Despite its fundamental importance, the formalization of quantum data encoding has never been addressed systematically so far. 
We formalize the concept of quantum data encoding, namely the format providing a representation of a data set through a quantum state, as a distinct abstract layer with respect to the associated data loading circuit.
We survey existing encoding methods and their respective strategies for classical-to-quantum exact and approximate data loading, for the quantum-to-classical extraction of information from states, and for quantum-to-quantum encoding conversion. Next, we show how major quantum algorithms find a natural interpretation in terms of data loading. For instance, the Quantum Fourier Transform is described as a quantum encoding converter, while the Quantum Amplitude Estimation as an extraction routine.
The new conceptual framework is exemplified by considering its application to quantum-based Monte Carlo simulations, thus showcasing the power of the proposed formalism for the description of complex quantum circuits.
Indeed, the approach clarifies the structure of complex quantum circuits and enables their efficient design.
}

\section{Introduction}\label{sec:intro}

\begin{figure}[t]
    \centering\small
\includegraphics{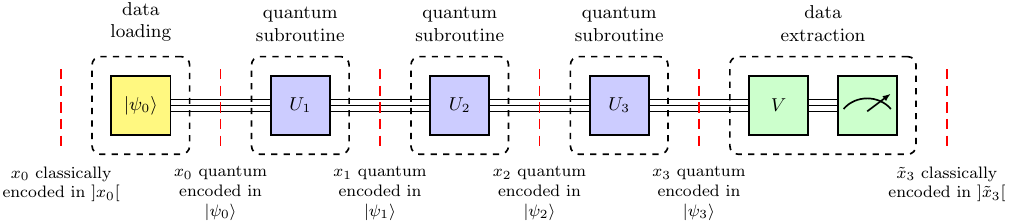}
    \caption{A quantum circuit embodied for the gate-model architecture can be interpreted as a quantum data loading procedure, followed by a collection of quantum subroutines, and finally by a data extraction process. A data loading routine takes as input information encoded in a classical structure, such as for instance a classical variable, an array, or a binary tree, and it produces a state that represents the same information in a given quantum encoding. A quantum subroutine takes an input state representing some information in a given encoding, and produces an output state representing new information in another given encoding. A data extraction routine takes an input state in a given encoding, and returns part of the information as an output register of bits, after a measurement process. Typically, a classical post-processing of multiple shots is needed to retrieve a significant amount of information from the quantum state.}
    \label{fig:summary}
\end{figure}

\begin{figure}[t]
    \centering\small
\includegraphics{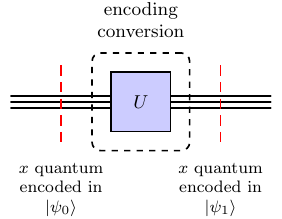}
    \caption{An encoding conversion routine is a special quantum subroutine that takes some information $x$ in input as a state in a given encoding, and returns the same information in output, as a state in a different encoding.}
    \label{fig:conversion}
\end{figure}

Gate model quantum computers \cite{QCladd2010quantum,QChuang2020superconducting,QCcong2022hardware,QCrotta2017quantum,ferraro2020all,QCde2023silicon,QCmanovitz2022trapped} process quantum algorithms based on sequences of one-qubit and two-qubits quantum logic gates. Such algorithms involve sequences of gates generally grouped into quantum subroutines \cite{heim2020quantum}. 
All quantum subroutines receive an input encoded by a quantum state in some specific form, and similarly provide an output in a suitably designed state. The representation of information through quantum states therefore plays a key role in the connection of quantum subroutines, thus conditioning the overall design of quantum algorithms. The encoding also determines the time and space required for data loading and for information extraction, thus affecting the performance of an algorithm.
Here, we formally classify the different ways to encode information in quantum states. We discuss the computational complexity of loading data in a given encoding from classical memories, of converting a quantum encoding into another one, and of retrieving classical information from quantum states into classical memories, respectively.

Despite a widespread awareness of the effects of data loading on runtimes, algorithmic literature still lacks homogeneity in terms of naming conventions, and the importance of subtle details of encodings is sometimes underestimated. The present work is therefore potentially beneficial for the design of efficient quantum algorithms. At the same time, we are developing a higher-level approach to quantum programming, that abstracts from the complexity of the implementations of subroutines, and describes said subroutines in terms of their input and output encodings.
A key motivation for our work is the establishment of a growing literature of complex quantum circuits resulting from the combination of underlying blocks (Fig.~\ref{fig:summary}). For example, quantum-based Monte Carlo simulations in finance~\cite{woerner_quantum_2019, egger_credit_2021, rebentrost_quantum_2018-1, ghosh_energy_2024} or physics~\cite{agliardi_quantum_2022} combine the loading of a random distribution with some form of function processing, to then extract output via Quantum Amplitude Estimation~\cite{maronese2023maximum}. More advanced examples~\cite{doriguello_quantum_2022} rely on quantum arithmetic~\cite{nelsen_introduction_1999}, namely the ability to perform arithmetic calculations operating on basis states, as we shall discuss later. In Ref.~\cite{agliardi_conditions_2024} a similar structure is leveraged, based on data loading, data processing, and data extraction, where the processing in this case is performed on amplitude-encoded states through the Quantum Hadamard Product~\cite{holmes_nonlinear_2021} and the swap test. The terminology and the examples above are clarified in the rest of the paper. For the moment, suffice it to say in all such cases, quantum encodings offer a helpful description of how subroutines operate on data.

In prior literature, efforts have already been made to collect and discuss common patterns in the design of quantum algorithms~\cite{feld_towards_2019, gilliam_foundational_2021}. Along this direction, M. Weigold \textit{et al.}~\cite{weigold_data_2020, weigold_expanding_2021, weigold_encoding_2021} have focused a greater part of their attention to data loading patterns, providing the most systematic and comprehensive work on the topic so far, at the best of our knowledge.

Compared to existing works, we introduce a marked distinction of two concepts: we call \textit{data encoding} the formal representation that a data set takes as a quantum state. By \textit{data loading} instead, we mean the state preparation routine that needs to be executed to physically clone data from a classical memory to a quantum register in a given encoding. Such distinction reflects the observation that information can be encoded in a quantum state not only through a loading procedure, but also through quantum processing, namely by a  calculations on the quantum computer. Additionally, the same data encoding can be obtained via alternative data loading procedures, all the more so if one considers also approximate loading techniques.
As a special case of subroutines, we introduce \textit{encoding converters} (Fig.~\ref{fig:conversion}), that allow to switch from one encoding to the other. Indeed, the ability to perform encoding conversion is essential in order to apply some specific algorithms or to obtain improved loading performances. We anticipate that the Quantum Fourier transform can be read under the light of an encoding converter. Literature is rich in works addressing individual and specific techniques for data loading or data conversion. We collect such contributions in the next Sections, connecting them in our conceptual framework.
We pairwise devote attention to \textit{data extraction}, namely the set of techniques that allow to retrieve classical information from a quantum state in a given encoding. We specifically interpret the Quantum Amplitude Estimation~\cite{brassard_quantum_1998} from this perspective.

The framework is applied to quantum-based Monte Carlo simulations, as an example of how our newly introduced conceptualization can be helpful in the design of nontrivial circuits.

In the next Section we introduce various quantum encoding methods, initially for a single data point, then for data sets (namely, collection of data points), and finally for multiple data sets. Section~\ref{sec:loading} contains methods for loading data, both exactly and approximately, and to convert one encoding into another. The complementary task of extracting data from a quantum state into a classical information is discussed in Section~\ref{sec:extraction}. The Section~\ref{sec:montecarlo} is devoted to the application of the new framework to a well-known quantum algorithm for Monte Carlo simulations. Finally, Section~\ref{sec:conclusions} contains the conclusions and the future outlook.

\section{Quantum encoding methods}
In this Section we present multiple forms in which classical information can be encoded through a quantum state. We start with the encoding of a single value in Subsec.~\ref{subsec:encoding-dp}, to then tackle entire data sets in~\ref{subsec:encoding-ds} and~\ref{subsec:encoding-qv-anticipation}, and finally multiple data sets in Subsec.~\ref{subsec:encoding-mds}. Afterwards, Subsec.~\ref{subsec:swap-dc}, \ref{subsec:amplitude-unifying}, and~\ref{subsec:beyond-pure} offer some remarks about the impact of encoding variances on quantum processing, on the conceptual importance of distinct encodings, and on the role of mixed states in encodings, respectively. Lastly, Subsec.~\ref{subsec:encoding-fn} is concerned with the encoding of functions, discussing how a quantum operator can be used to represent a classical function, by acting on data in a given encoding.

\subsection{Encoding of a data point}\label{subsec:encoding-dp}
There are three main encoding methods of a single data point, namely the basis, angle and Fourier encoding respectively.

\paragraph{Basis encoding.} Let $x$ be a integer value in $\{0, ..., 2^m-1\}$, and write its binary decomposition as $x=\sum_{j=0}^{n-1} x_j 2^j$ with $x_j \in \{0,1\}$ for all $j$. Similar to the classical representation, the number can be simply encoded by the state
\begin{equation}\label{eq:basis-encoding}
\ket{\psi} := \bigotimes_{j=n-1}^{0} \ket{x_j} = \ket{x_{n-1}} \cdots \ket{x_0}
\end{equation}
of $m$ qubits, namely $\ket{\psi} \in \otimes^m \mathcal H$, where $\mathcal H = \mathbb C^2$ is the single-qubit Hilbert space. In Eq.~\eqref{eq:basis-encoding}, $\ket{0}$ and $\ket{1}$ are two conventional orthogonal states of a qubit, forming the so-called \textit{computational basis of the qubit}. The state~$\ket{\psi}$ is known as the \textit{basis encoding} of $x$ and compactly denoted by $\ket{x}$.

\paragraph{Angle encoding (or qubit encoding).}
Given a real number $\theta \in \left[0, \frac{\pi}{2}\right]$, its \textit{angle encoding} is the state
$$\ket{\psi} := \mathtt{R_Y}(2 \theta) \ket{0} =  \cos \theta \ket{0} + \sin \theta \ket{1}$$
on a single qubit.

The angle encoding is clearly efficient in terms of space, as it only requires one qubit for a data point, compared to $m$ qubits needed by the basis encoding. Additionally, assuming a perfect hardware able to control qubit rotations with arbitrary precision, the angle encoding can store an arbitrary real value from the domain $\left[0, \frac{\pi}{2}\right]$ of infinite cardinality, while the basis encoding on $m$ qubits contains a value from a range of just~$2^m$.

\paragraph{Fourier encoding.}
Let $x$ be an integer value in $\{0, ..., 2^m-1\}$. We decompose $x$ into its binary digits $x = \sum_{j=0}^{m-1} x_j 2^j$. We also denote the corresponding binary fraction as
\begin{equation}\label{eq:binfrac}
    \overline{.x_0 x_1 \dots x_{m-1}} := \sum_{j=0}^{m-1} x_j 2^{-j-1}.
\end{equation}
We introduce the term \textit{Fourier encoding} of $x$ or equivalently of $\overline{.x_0 x_1 \dots x_{m-1}}$, for the state
$$\ket{\psi} := \frac{1}{\sqrt{2^m}} \bigotimes_{j=0}^{m-1} \left[ \ket{0} + e^{2 i \pi \, \overline{.x_j \dots x_{m-1}}} \ket{1} \right] = \frac{1}{\sqrt{2^m}} \bigotimes_{j=0}^{m-1} \mathtt P(2 \pi \, \overline{.x_j \dots x_{m-1}}) \ket{0}$$
on $m$ qubits, where $\mathtt P$ is the Phase gate, $\mathtt P(\theta) = \begin{pmatrix} 1 & 0 \\ 0 & e^{i \theta} \end{pmatrix}$.

The importance of the encoding lies in the relationship with the Quantum Fourier Transform, that also motivates its name, as shown in Subsection~\ref{subsec:load-point}.

Given the above basic definitions, the characteristic of the different encodings become manifest when dealing with multiple data, as we discuss in the remainder of the Section.

\subsection{Encoding of a data set}\label{subsec:encoding-ds}
In this Subsection, we extend the definition of encodings to \textit{collections} of values. We will refer to the entire collection as \textit{data set}, and to its items as \textit{data points}.

\paragraph{Multi-register encoding.}
Let $[x_i]_{i=0}^{N-1}$ be a collection of $N$ integer values, each in $\{0, ..., 2^m-1\}$. By mimicking classical arrays, one can store each data point in a separate register in the basis encoding, thus constructing the state
\begin{equation*}
    \ket{\psi} := \bigotimes_{i=0}^{N-1} \ket{x_i}
\end{equation*}
on $mN$ qubits, namely the \textit{multi-register encoding} of the data set.

This encoding is very expensive in terms of qubits, as it does not exploit the quantum superposition. On the positive side, though, one can easily define quantum arithmetic operations: given $\ket{x_0} \ket{x_1} \ket{0}$, there exist unitary operators $U$ such that 
\begin{equation}\label{eq:arithmetic-operators}
    U \ket{x_0} \ket{x_1} \ket{0} = \ket{x_0} \ket{x_1} \ket{x_{\mathrm{out}}},
\end{equation}
where $x_{\mathrm{out}}$ may be the modular sum, modular product, maximum, minimum, etc, according to the definition of $U$~\cite{nielsen_quantum_2010}. At this stage such operators do not appear very useful compared to the classical counterparts, but we discuss them again later, when writing of entangled encodings in Subsection~\ref{subsec:encoding-mds}.

\paragraph{Equally-weighted encoding (also called digital encoding or quantum associative memory QuAM).}
Let $[x_i]_{i=0}^{N-1}$ be a collection of $N$ integer values, each in $\{0, ..., 2^m-1\}$. Their \textit{equally-weighted encoding} is the superposition of the basis encodings for each data point, represented by the state
$$\ket{\psi}:= \frac{1}{\sqrt{N}} \sum_{i=0}^{N-1} \ket{x_i}$$
on $m$ qubits.

A particular case is the full uniform superposition of all qubits, obtained when $N=2^m$ and $[x_i]_i = [0, ..., 2^m-1]$. Many algorithms, including Shor's, start in uniform superposition.

\paragraph{Angle encoding.}
Given data points $[\theta]_{i=0}^{N-1}$, each lying in $\left[0, \frac{\pi}{2}\right]$, the tensor product of their angle encodings is referred to with same term \textit{angle encoding}. The resulting state is then
$$\ket{\psi} := \bigotimes_{i=0}^{N-1} \left[ \cos \theta_i \ket{0} + \sin \theta_i \ket{1} \right]$$
on $N$ qubits.

Angle encodings are widely used in Quantum Machine Learning~\cite{larose_robust_2020} and are natively supported in dedicated software libraries like Pennylane~\cite{bergholm_pennylane_2022}.

\paragraph{Amplitude encoding (also called analog encoding).}
Let $[a_i]_{i=0}^{N-1}$ be a collection of $N \leq 2^n$ complex values constrained by $\sum_i \abs{a_i}^2 = 1$. Without loss of generality, we can assume $N = 2^n$ by completing $[a_i]_i$ with zeros. Their \textit{amplitude encoding} is the state
$$\ket{\psi} := \sum_{i=0}^{N-1} a_i \ket{i}$$
on $n$ qubits.

Elaboration of data encoded by amplitudes is constrained to the possibilities offered by qubit rotations. For instance, the calculation of piecewise linear functions by means of local sinusoidal approximations~\cite{woerner_quantum_2019,gacon_quantum-enhanced_2020} has become a key example in this context. A more sophisticated example is the Quantum Hadamard Product used to perform nonlinear operations to quantum information in amplitude encoding through measurements~\cite{holmes_nonlinear_2021}. The Quantum Amplitude Estimation algorithm~\cite{brassard_quantum_1998} and its variants, introduced in Section~\ref{sec:extraction}, can be seen as a way to efficiently perform Monte Carlo simulations (see Section~\ref{sec:montecarlo}), assuming input data are encoded by amplitudes.

Let us highlight that given a discrete random variable $X$ valued in $\{0, ..., N-1\}$, with densities $p_i=\mathbb P(X=i)$, for $i=0, ..., N-1$, it is very natural to represent it in amplitude encoding by the state
$$\ket{\psi} := \sum_{i=0}^{N-1} \sqrt{p_i} \ket{i}:$$
indeed, the probability of measuring an output $i$ is exactly $p_i$.

\paragraph{Divide \& conquer encoding (D\&C encoding).}
Let $[a_i]_{i=0}^{N-1}$ be a collection of $N=2^n$ complex values indexed over $i=0, ..., N-1$ and constrained by $\sum_i \abs{a_i}^2 = 1$. Their \textit{D\&C encoding} is the state
$$\ket{\psi} := \sum_{i=0}^{N-1} a_i \ket{i} \ket{\psi_i}$$
defined on two registers, the first of size $n$ and the second of size $N$, where $[ \ket{\psi_i} ]_i$ is a suitable collection of auxiliary states~\cite{araujo2021divide} of $N$ qubits.

The reason of interest for this encoding is that the associated loading unitary can be executed in $O(\log_2^2 N)$ depth (under the additional hypothesis that classical data are stored in a suitable binary tree) compared to $O(N)$ of the general data preparation in amplitude encoding. The D\&C can be therefore seen as a loading-efficient, but qubit-intensive variant of the amplitude encoding.

Beyond being interesting \textit{per se}, we specifically introduced D\&C to prepare the discussion in Subsection~\ref{subsec:swap-dc}, where we show that transitioning from the amplitude encoding to the D\&C encoding can have huge impacts on processing algorithms, despite the apparently small difference between the two encoding formats.

\begin{figure}
    \centering \small
    \input{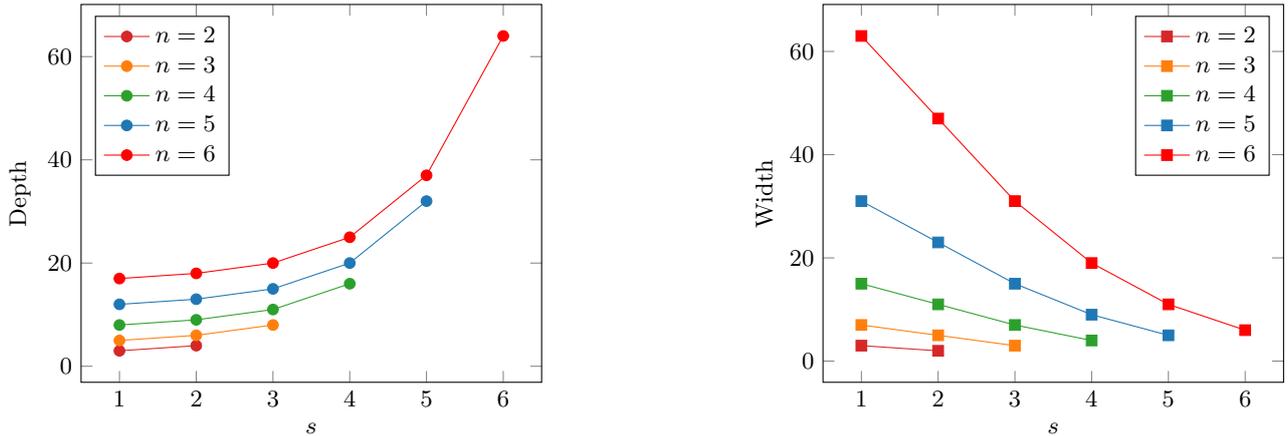}
    \caption{Width and depth of the loading circuit for the bidirectional encoding~\cite{araujo2021divide,araujo_configurable_2022}, as a function of the split level $s$ and number of qubits $n$.}\label{fig:bidir}
\end{figure}

\paragraph{Bidirectional encoding.}
In the attempt to balance circuit width and depth, it is possible to define intermediate stages between the amplitude encoding, that is qubit-efficient, and the D\&C, that is loading-efficient. Specifically, taken a normalized vector $[a_i]_{i=0}^{N-1}$ of size $N=2^n$, the \textit{bidirectional encoding}~\cite{araujo_configurable_2022} is parametrized over a so-called \textit{split level} ranging in $1,...,n$, and gives rise to a width $O((s+1)N 2^{-s})$ and a loading depth of $O(2^s + n - s^2)$, thus retrieving the performance of amplitude and D\&C encoding respectively, in the two extreme cases (see Fig.~\ref{fig:bidir}). Representations are still in the form
$$\ket{\psi} := \sum_{i=0}^{N-1} a_i \ket{i} \ket{\psi_i}$$
but the auxiliary states $[\ket{\psi_i}]_i$ have a different size and therefore also a different content. Still, literature lacks a full characterization of which algorithms designed for amplitude encoding can be extended to the bidirectional encoding.

\paragraph{Generalized amplitude encoding.}
We use the term \textit{generalized amplitude encoding} to refer to any encoding in the form
$$\ket{\psi} := \sum_{i=0}^{N-1} a_i \ket{i} \ket{\phi_i}$$
where $[\ket{\phi_i}]$ are auxiliary states in given ancillary registers, of any size or form, thus including the amplitude encoding, the D\&C encoding, and the bidirectional encoding introduced above, among others. The QAE properly works on such states with $N=2$, as we discuss in Section~\ref{sec:extraction}.

\paragraph{Embeddings.}
Other ways of embedding classical information into a quantum circuit were introduced, particularly in the context of Quantum Machine Learning (QML)~\cite{perez-salinas_data_2020, schuld_effect_2021}. In this case, data are loaded with the aim of \textit{influencing} the behavior of a quantum neural network. The state preparation is often interleaved with the processing, so that there is no single moment in which loaded data can be accessed through measurement. We use the term \textit{embeddings} to refer to such situations, and we deliberately exclude them from our analysis, as they do not contribute to our objective of modular algorithm design.

\subsection{Encoding with mappings}\label{subsec:encoding-qv-anticipation}
In computer science, the representation of a variable determines the set of values allowed for the variable. We call \textit{domain} said set of values.

So far, we have discussed \textit{native} encodings, in the sense that input data satisfied the requirements that their encodings naturally demanded for their domains (non-negative integer numbers in the basis encoding, real numbers in $[0, \frac{\pi}{2}]$ for the angle encoding, normalized vectors for the amplitude encoding, etc.). More in general, we can think of a classical preprocessing that transforms the input data in the natural domain before executing the quantum circuits, and an inverted postprocessing that transforms back to the original domain.

In the case of amplitude encoding, it is clear how to treat a non-normalized vector: the input is normalized before loading it, and the output is appropriately de-normalized at the end of the quantum workload. This transformation has two effects that must be taken in careful consideration: on one side, it implies a classical pre- and post-processing cost of $O(N)$, that in the extreme cases may undermine the quantum speedup, and on the other side, may introduce an error propagation that again shadows the benefits of quantum algorithms, for instance when the output needs to be rescaled superlinearly with the input normalization factors (see e.g. Ref.~\cite{holmes_nonlinear_2021}).

\begin{figure}
    \centering\small
    \includegraphics{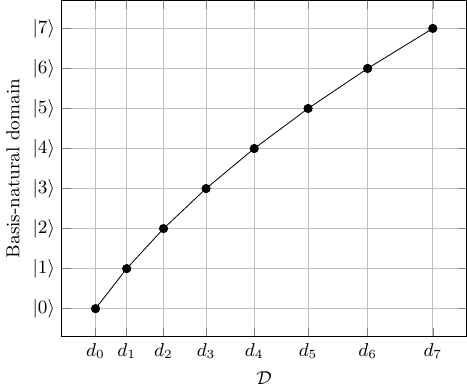}
    \caption{An example of a bijection $g$ realizing an encoding: here, $\ket{j}$ is used to represent the value $d_j=g^{-1}(j)$.}\label{fig:g-encoding}
\end{figure}

Similar considerations hold for the basis encoding: given a bijection $g$ between the basis-natural domain $\{0, ..., 2^m-1\}$ and any other domain $\mathcal D$, one can represent elements in $\mathcal D$ by the basis encoding, see Fig.~\ref{fig:g-encoding}. The effectiveness of the representation obviously depends on the ability to make manipulations in the quantum space that correspond to useful computations in the original space of $\mathcal D$, see Remark~\ref{rem:montecarlo-with-mapping}. Examples of said technique include fixed point~\cite{yang_quantum_2022} and floating point~\cite{haener_quantum_2018, seidel_efficient_2022, agliardi_floating_2023} representations of real numbers on quantum computers.

\subsection{Encoding of two data sets}\label{subsec:encoding-mds}
Earlier in this Section, we transitioned from encodings of single data points to encodings of whole data sets, showing that few encodings of a point give rise to a multiplicity of encodings for data sets. Similarly, it may not be surprising that many options are available to combine several data sets. The choice of an encoding, among the different possibilities, is related to the role of each data set in the computation. At the same time, the chosen encoding implies meaningful relationships among the data sets themselves.

\paragraph{Amplitude encoding reinterpreted.} Under the lens of the mappings, the amplitude encoding $\sum_i a_i \ket{i}$ is already a representation of two data sets: the one of amplitudes, and the one of indexes. As expected, amplitudes and respective indices are tightly connected in the amplitude encoding. In this Subsection we shall read this fact as a fully entangled encoding among two data sets.

\paragraph{Independent encodings.} Given two data sets, they can be independently encoded by a given encoding format into the states $\ket{\psi_1}$ and $\ket{\psi_2}$, so that their tensor product $\ket{\psi_1 \psi_2}$ is an encoding for the couple. Independent encodings due their name to the fact that the measurement of the respective register give rise to independent random variables \cite{agliardi_floating_2023}.

An example of arithmetic operator acting on independent amplitude encodings is the Quantum Hadamard Product~\cite{holmes_nonlinear_2021}, which calculates the component-wise product of two vectors represented by independent amplitude encodings. It should be remarked though that the product, seen in such terms, is subject to a success probability depending on the vector norms.

\paragraph{Fully entangled encodings.} The opposite case is that of fully entangled encodings. In a probabilistic interpretation, it corresponds to variables defined on the same underlying events: for instance, starting from amplitude encodings,
$$\ket{\psi} := \sum_i \sqrt{p_i} \ket{x_i} \ket{y_i} $$
represents the fact that $X$ takes the value $x_i$ \textit{exactly when} $Y$ takes the value $y_i$. The concept naturally extends to complex coefficients.

Arithmetically, this is useful in combination with the operators introduced with Eq.~\eqref{eq:arithmetic-operators}: by exploiting linearity of quantum operators, they can work on superposed inputs in \textit{quantum parallel}. For example, suppose two vectors $x_i$ and $y_i$ are encoded by the weighted states
$$\ket{\psi} := \sum_i a_i \ket{x_i} \ket{y_i} \ket{0}$$
where $a_i$ are any weights, for instance $a_i = N^{-1/2}$ for the equally weighted states. Then the operator $U$ that performs addition, in the sense of Eq.~\eqref{eq:arithmetic-operators}, in this context will output
$$U \ket{\psi} = \sum_i a_i U \ket{x_i} \ket{y_i} \ket{0} = \sum_i a_i \ket{x_i} \ket{y_i} \ket{x_i+ y_i} $$
in a single application~\cite{agliardi_floating_2023}.

\paragraph{qRAM encoding.} As a notable case of fully entangled encodings, we have the qRAM encoding, that represents two data sets, one of amplitudes $[ a_i ]_i$ and one of integers $[ x_i ]_i$ indexed over the same $i$, in the form
$$\sum_i a_i \ket{x_i} \ket{i}.$$
Its importance lies in the conjectured ability to efficiently load data in such format through appropriate devices, as discussed in the next Section.

\paragraph{Partially entangled encodings.} More broadly speaking, given two registers, in the probabilistic view, their states can be interpreted as
$$\ket{\psi} := \sum_{ij} \sqrt{p_{ij}} \ket{x_i} \ket{y_j}, $$
when coefficients are real and nonnegative, so that we are dealing with two discrete random variables with given joint probabilities. Again, this intuitive view extends to the complex coefficients.

\subsection{Effects of the encoding variants on the computation}\label{subsec:swap-dc}
\begin{figure}
\centering \small \hspace{0cm}
\includegraphics{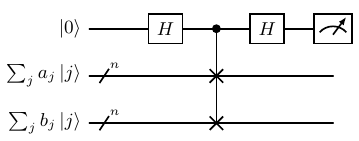}
\caption{The circuit for the swap test. The probability of the measurement to be $0$ is $\frac{1}{2} + \frac{1}{2} \abs{ \sum_j a_j b_j }^2$. Here, $H$ is the Hadamard gate.}
\label{fig:swap}
\end{figure}

When proposing the D\&C encoding, the authors also highlight~\cite{araujo2021divide} that, despite the seemingly small difference from the amplitude encoding, not every algorithm designed for amplitude encoding can be applied to data in D\&C. Specifically, this is shown for the swap test, often adopted for the calculation of the inner product of vectors (Fig.~\ref{fig:swap}). The authors finally demonstrate that the D\&C encoding can in turn be modified, by adding $n$ qubits and without asymptotic depth overhead in the loading procedure, into another encoding that we call D\&C-orthonormal, which allows for the application of the swap test. Interestingly, while the usual swap test in amplitude encoding provides an estimation of $\abs{ \sum_j a_j b_j }$, in the D\&C-orth encoding it outputs $\sum_j \abs{a_j}^2 \abs{b_j}^2$. The same orthonormal modification is also applicable for the bidirectional encoding, with the same effect on the output of the swap test~\cite{agliardi_conditions_2024}.

Beyond the specific example of the swap test, what we want to emphasize here is that an apparently minor change in the encoding, namely the entanglement of the exact same information with additional side registers, can turn a processing algorithm to be inapplicable or to provide a different result.

\subsection{A unified view?}\label{subsec:amplitude-unifying}

Given that every pure state writes in the form
$$\ket{\psi} = \sum_{i=0}^{N-1} a_i \ket{i},$$
with some complex coefficients $a_i$, one may shortly argue that the amplitude encoding is the universal representation, and there is no need to introduce other encodings. This simplistic view hides important facts. First, in terms of data loading, the amplitude encoding requires $O(N)$ depth, while particular structures can lead to sparsity and perform more efficiently, as we shall see in Section~\ref{sec:loading}. Second, besides information encoding, it is of paramount importance to track the role that a data set plays in the computation, for the purpose of algorithmic design.

Indeed, from a computational point of view, a standardized lexicon for encodings is fundamental to connect subroutines, since each subroutine expects information in a given form, as we stressed with the example of the swap test.
Sometimes, conversion (see Fig.~\ref{fig:conversion}) between encodings is required to combine multiple subroutines together. For instance, HHL (named after Aram Harrow, Avinatan Hassidim, and Seth Lloyd~\cite{harrow_quantum_2009}) takes its input in the amplitude encoding, but it relies on the equally-weighted encoding of a matrix eigenvalues at an intermediate stage of the computation~\cite{mitarai_quantum_2019}. Other examples of algorithms based on multiple encodings are constituted by the quantum Metropolis sampling~\cite{mitarai_quantum_2019} and Variational Quantum Algorithms (VQA, see for instance Ref.~\cite{rebentrost_quantum_2018-1}).

On the other hand, from a physical point of view, the same quantum state can be \textit{interpreted} according to diverse encodings, each suited for a specific conceptualization. For instance, considering the state $\sum_i a_i \ket{x_i} \ket{y_i} \ket{0}$, it may be convenient to consider $\ket{x_i, y_i, 0}$ as a single register for the purpose of the data loading, and then as three separate register in a computation step, for instance to apply an arithmetic circuit, say the sum. Intended as the lens through which the state is read, the encoding is contextual to the algorithm, and such fact justifies the lack of organicity in treatment of the topic across different sources.

\subsection{Beyond pure states}\label{subsec:beyond-pure}
In this work we always deal with pure states, so that the notations introduced so far are enough for our purposes. Nevertheless, we should at least mention two contexts in which this is not the case.

First of all, quantum algorithms are usually concerned with pure states (with some notable exceptions, see Ref.~\cite{bermejo_variational_2022}), but the broader field of quantum computing is not. Indeed, perfect system isolation and perfect qubit control is assumed in algorithm design, so that pure states describe sufficiently well algorithms at a logical level, which is the core interest in this work. Mixed states acquire great relevance instead when applying error corrections to hardware-executed algorithms.
Throughout the paper, since we work with pure states, we can represent them as state vectors $\ket\psi$. More generally, a (mixed) state is represented by a \textit{density operator}~\cite{nielsen_quantum_2010}, associated to a density matrix $\rho$ of size $2^n \times 2^n$, where $n$ is the number of qubits. The density operator must be positive semi-definite and Hermitian (namely, self-adjoint), and its trace must be $1$. A state is pure if the density matrix is the outer product of a vector with itself $\rho = \ketbra{\psi}{\psi}$, or equivalently if $\rho = \rho^2$. The trace of $\rho^2$ is called purity, and the purity equals $1$ when and only when the state is pure. Mixed states represent ensembles of pure states, namely they can be decomposed as a collection of pure states, each associated to its own probability: $\rho = \sum_j p_j \ketbra{\psi_j}{\psi_j}$. It is clear then that mixed states model the effect of measurements on (pure) states: indeed, we know that, by measuring, the initial state gets projected to some eigenstate $\ket{\psi_j}$ of the measurement operator, associated to the measured eigenvalue, and we know that this happens with a given probability $p_j$ determined by the interaction between the initial state and the measurement operator. The purpose of mixed states is exactly to represent the outcome of processed and measured input states, where measurement can occur for computational purposes, or be the undesired effect of decoherence, thus explaining the relation with noise modeling.

\begin{figure}
    \centering \small \hspace{0cm}
    \input{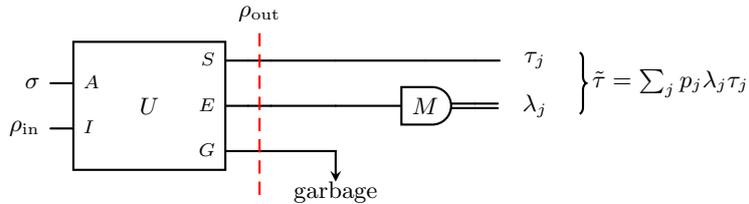}
    \caption{The weighted state $\tilde \tau$ is the partial output of a quantum routine that takes as inputs an input state $\rho_{\mathrm{in}}$ in a register $I$ and an ancilla state $\sigma$ in a register $A$, then applies a unitary $U$ and a measurement $M$ to the environment qubits $E$, and finally discards qubits in register $G$. While $\rho_{\mathrm{out}}$ is a physical state, the weighted state $\tilde \tau$ in general is not, as an effect of discarding $G$.}
    \label{fig:weighted-states}
\end{figure}

Secondly, it is worth mentioning that an increasing number of algorithms exploits mid-circuit measurements as a way to perform non unitary, non reversible quantum computation~\cite{dynamic_circuits}, also fostered by the recently introduced opportunity to run such capabilities of `dynamic circuits' in commercially available hardware~\cite{noauthor_get_2020, noauthor_quantum_2021}. Additionally, the need for nonlinear transformation suggests the discard of some qubits after performing computations in wider Hilbert spaces. In this framework, \textit{weighted states} were proposed~\cite{holmes_nonlinear_2021} as a formalism. Despite being more complex than the usual bra-ket notation, this formalism will likely be useful in algorithms and applications making extensive usage of mid-measurements.
More in detail, a weighted state is derived from following observation, as depicted in Fig.~\ref{fig:weighted-states}. A quantum algorithm takes an input state $\rho_{\mathrm{in}}$ and some ancilla qubits in a state $\sigma$, applies a unitary operation $U$, and outputs a state $\rho_{\mathrm{out}}$. Now, the outcomes of $\rho_{\mathrm{out}}$ are deeply constrained by the linearity of quantum operators, so that it is impossible to obtain a nonlinear transformation out of this schema. Therefore, it is a common practice to prepare $\rho_{\mathrm{out}}$ in a wider space, and to look only at a \textit{portion} of the state, which can then show nonlinear behaviors. As represented in the Figure, the full state undergoes a unitary transformation $U$, followed by the measurement of some qubits through an operator $M$, and the discard of some others. At the output of $U$, qubit registers are labelled as the system $S$ (the portion of interest), the environment $E$ (subject to measurement) and the garbage $G$ (subject to discard). If we define $\tilde \tau := \trace_{EG} \left( \rho_{\mathrm{out}} \left( I_S \otimes M_E \otimes I_G \right)\right)$, we get a so-called \textit{weighted state}: an object that may not correspond to a physical state, but generalizes it. In particular, it behaves as a state whenever an operator is applied to the system register: by linearity, indeed, $\trace_{SEG} [\tilde \tau O] = \trace_{SEG} \left( \rho_{\mathrm{out}} \left( O \otimes M_E \otimes I_G \right)\right)$, for any linear operator $O$ applied to the system qubits. A more detailed description can be found in the original paper~\cite{holmes_nonlinear_2021}.

\subsection{Encoding of functions}\label{subsec:encoding-fn}
Given a classical function $f$, one needs to define a unitary operator $O_f$ that replicates the behavior of $f$ acting on quantum registers. Clearly, the circuit that executes $O_f$ depends on the encodings that input and output states of $O_f$ are supposed to have. In this Subsection, we list few function encodings that are crucial in quantum algorithms.

\paragraph{Digital encoding of $f$.} A digital encoding of $f$ is a function encoding that takes input in basis encoding and provides output in basis encoding. More precisely, let $f$ be a function acting on one integer in $\{ 0, ..., n_{\mathrm{in}} -1 \}$ and valued in $\{0, ..., n_{\mathrm{out}}-1\}$. Then $O_f$ acting on $n_{\mathrm{in}}+n_{\mathrm{out}}$ qubits is called a \textit{digital encoding} of $f$ if
$$O_f \ket{x} \ket{y} = \ket{x} \ket{f(x) + y \; \mathrm{mod} \, n_{\mathrm{out}}} $$
for any $x$ in the domain, $y$ in the codomain. In the typical usage, the second register is initialized to $y=0$, so that after the gate application, it contains exactly the desired information. The definition trivially extends to functions that take vectors as inputs, or provide vectors as outputs. For instance, the operations of modular sum, modular product, minimum, maximum that we introduced with Eq.~\eqref{eq:arithmetic-operators}, are the digital encoding of the respective functions $f_{\mathrm{sum}}(x_1,x_2) = x_1+x_2 \mod n_{\mathrm{out}}$, $f_{\mathrm{prod}}(x_1,x_2) = x_1 x_2 \mod n_{\mathrm{out}}$, $f_{\mathrm{min}}f(x_1,x_2) = \min \{ x_1,x_2 \}$, $f_{\mathrm{max}}(x_1,x_2) = \max \{ x_1,x_2 \}$.

\paragraph{Amplitude encoding of $f$.} The amplitude encoding of $f$ is a function encoding that takes input in basis encoding and provides output in angle encoding. More precisely, let $f$ be a function acting on one integer in $\{ 0, ..., n_{\mathrm{in}} -1 \}$ and valued in $[0, 1]$. Then $R_f$ acting on $n_{\mathrm{in}}+1$ qubits is called an \textit{amplitude encoding} of $f$ if
\begin{equation}\label{eq:fn-ampl-encoding}
R_f \ket{x} \ket{0} = \sqrt{1-f(x)} \ket{x} \ket{0} + \sqrt{f(x)} \ket{x} \ket{1}
\end{equation}
for any $x$ in the domain. Notice that the output is an angle encoding of $\arcsin f$, and not of $f$. The definition extends to functions that take vectors as inputs.

\begin{remark}[Quantum parallel execution]
Notice that all function encodings that take inputs in the basis states, including the digital encoding and the amplitude encoding here defined, allow for \textit{quantum parallel}, as we introduced in Subsection~\ref{subsec:encoding-mds}: given an input data set in the form $\sum_i a_i \ket{x_i}$, any digital encoding operator $O_f$ is able to evaluate the underlying function $f$ on the whole input in a single application, in the following sense: $O_f \sum_i a_i \ket{x_i} \ket{0} = \sum_i a_i \ket{x_i} \ket{f(x_i)}$. The same holds for an amplitude encoding operator $R_f$: $R_f \sum_i a_i \ket{x_i} \ket{0} = \sum_i a_i \left[ \sqrt{1-f(x_i)} \ket{x_i} \ket{0} + \sqrt{f(x_i)} \ket{x_i} \ket{1} \right]$. In both cases, this is a simple application of the linearity of quantum operators.
\end{remark}

\section{Data loading and encoding conversion}\label{sec:loading}
Loading classical data into a quantum computer in amplitude encoding is a critical task~\cite{grover_synthesis_2000,grover_creating_2002,mitarai_quantum_2019,sanders_black-box_2019} which jeopardizes the advantage of some core quantum processing algorithms~\cite{aaronson_read_2015} of many algorithms that are already classically linear in time, and may translate in sublinear scales. Examples include HHL (named after Aram Harrow, Avinatan Hassidim and Seth Lloyd~\cite{harrow_quantum_2009}) for the resolution of linear systems of equations, and the Quantum Fourier Transform (QFT)~\cite{coppersmith_approximate_2002} that is a quantum version of the fast discrete Fourier transform. Discarding the data loading cost, these techniques would have a logarithmic scaling in the input size, and therefore get an exponential speedup against the classical counterparts, that gets nullified by the loading unitary.
Consequently, besides the exploitation of native quantum data~\cite{le_flexible_2011}, in more recent times the idea of efficiently loading data \textit{approximately} has been discussed in literature. 

In this Section we discuss both exact and approximate techniques.

\subsection{Exact data loading of data points and encoding conversion}\label{subsec:load-point}
Given the definitions of basis encoding, angle encoding and Fourier encoding in Section~\ref{subsec:encoding-dp}, loading a single data point in such encodings is straight-forward. The problem becomes non-trivial when dealing with data sets instead of data points: the present Subsection is devoted to the topic. First, let us introduce the Quantum Fourier Transform and interpret it here as a tool for converting the basis encoding into the Fourier encoding and vice versa.

\begin{figure}
\centering \small \hspace{0cm}
\includegraphics{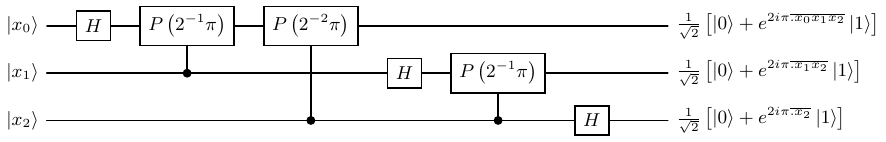}
\caption{The circuit of the Quantum Fourier Transform on 3 qubits, and its effect on a basis state $\ket{x_0 x_1 x_2}$.}
\label{fig:qft}
\end{figure}

\paragraph{The Quantum Fourier Transform.} The Quantum Fourier Transform (QFT)~\cite{nielsen_quantum_2010, j_quantum_2020} is an important tool in quantum information processing. It is a unitary operator that can be built through the circuit shown in Fig.~\ref{fig:qft}. Given an integer number $x$ in $\{ 0, ..., 2^m-1 \}$, the $m$-qubit QFT acts as
$$QFT \ket{x} = \frac{1}{\sqrt{2^m} } \sum_{j=0}^{m-1} e^{\frac{2 i \pi}{2^m}xj} \ket{j},$$
so that the formal analogy with the classical Discrete Fourier Transform is manifest~\cite{j_quantum_2020}. Here we emphasize that the QFT converts the basis encoding into the Fourier encoding: in the notations of Eq.~\eqref{eq:binfrac},
$$QFT \ket{x} = \frac{1}{\sqrt{2^m}} \bigotimes_{j=0}^{m-1} \left[ \ket{0} + e^{2 i \pi \overline{.x_j \dots x_{m-1}}} \ket{1} \right] = \frac{1}{\sqrt{2^m}} \bigotimes_{j=0}^{m-1} P(2 i \pi \overline{.x_j \dots x_{m-1}}) \ket{0}.$$

\subsection{Exact data loading of data sets and encoding conversion}\label{subsec:exact}
\begin{figure}[t]
    \centering
    \footnotesize
    \tikzstyle{classical} = [rectangle, rounded corners, minimum width=2.5cm, minimum height=1cm, text centered, text width=2.5cm, draw=black, fill=green!10]
    \tikzstyle{quantum} = [rectangle, rounded corners, minimum width=2.5cm, minimum height=1cm, text centered, text width=2.5cm, draw=black, fill=blue!10, node distance=3mm]
    \tikzstyle{process} = [rectangle, minimum width=2cm, minimum height=1cm, text centered, text width=2cm, draw=black, fill=orange!20]

    \includegraphics{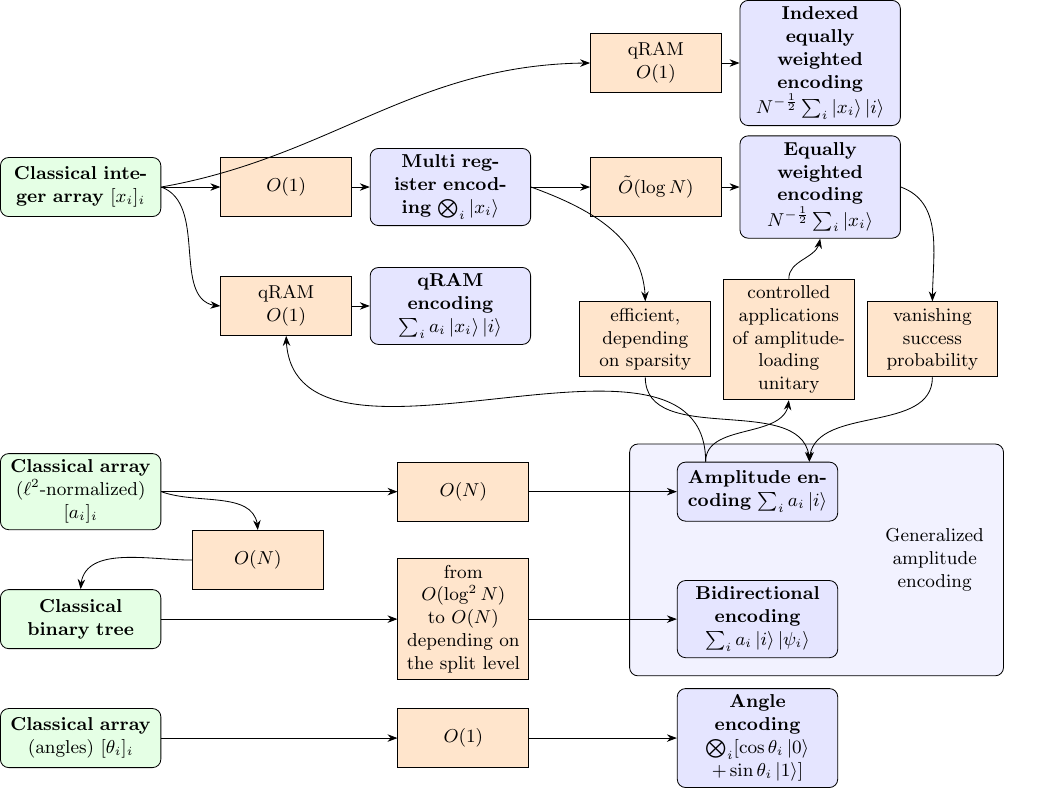}
    \caption{A summary of the different encodings (in green, classical, and in blue, quantum), and associated loading or conversion techniques (in orange). Big-O notations refer to circuit depths. The symbol $\tilde O$ is used when $\poly \log \log (\cdot)$ factors are discarded. `qRAM' marks the qRAM availability assumption. All the details are contained in Section~\ref{subsec:exact}.}
    \label{fig:loading}
\end{figure}

In this Subsection we deal with exact data loading. It will soon be clear that many times data in loaded in a temporary encoding and then converted, for efficiency. As a consequence, it appears natural to jointly treat loading and conversion. Fig.~\ref{fig:loading} summarizes the main techniques described in this Subsection.

\paragraph{Multi-register encoding and angle encoding.} Since these encodings derive from the tensor products of one-state preparations, their depth is $O(1)$ when the number of data points $N$ grows. On the downside, they are extremely qubit-intensive.

\paragraph{Equally-weighted encoding.} An equally-weighted encoding can be prepared efficiently by loading first data in the multi-register encoding, in $N=2^n$ qubits for a data set of size $N$, and then swapping until the information is accumulated in the desired subset of $n$ qubits~\cite{cortese_loading_2018}. The overall procedure requires $O(n \log n)$ depth, so it is very time-efficient, but the intermediate multi-register encoding implies a high qubit overhead. It should be remarked that the additional qubits can be left \textit{disentangled} from the core qubits at the end of the loading procedure: therefore the encoding is genuinely the equally-weighted.

\paragraph{Amplitude encoding.} We already mentioned that the amplitude encoding in the general case requires $O(N)$ depth and CNOTs. A loading technique can be found in Ref.~\cite{shende_synthesis_2006}, exploiting the inverse of a so-called \textit{quantum multiplexor}.

\paragraph{Multi-register to amplitude.} It is possible to translate from a multi-register encoding into an amplitude encoding, by means of a time-efficient protocol for non-sparse data~\cite{ashhab_quantum_2022}. Once again, passing through a multi-register encoding is inefficient in terms of qubits.

As a remark, above we focused on the \textit{quantum cost} (i.e., depth) of data loading. Since current access to classically stored information is performed sequentially, though, no exact loading algorithm can run faster than $O(N)$, if we consider it end-to-end, including the classical processing time needed to prepare the circuit. Such constraint could be overcome only with an efficient Quantum RAM (or qRAM), described below, or alternatively by feeding the circuit by native quantum data, collected for instance from a quantum sensor.

\paragraph{The qRAM.} By this term, we refer an operator $U$ performing the query access in quantum parallel, in constant time. In practical terms, this means that given a data set $[ x_i ]_i$ in a classical memory, $U$ performs as follows:
\begin{equation}\label{eq:qgan-encoding}
    U : \sum_i a_i \ket{i} \ket{0} \mapsto \sum_i a_i \ket{i} \ket{x_i}.
\end{equation}
The operator is commonly associated to a physical device allowing such operation. Experimental demonstrations of said devices are proposed for instance in Refs.~\cite{giovannetti_architectures_2008,hann_hardware-efficient_2019}, but they are not available in commercial quantum systems at scale, at the best of our knowledge.

Eq.~\eqref{eq:qgan-encoding} shows that qRAM devices are able to load information in the qRAM encoding, \textit{assuming that} the state $\sum_i a_i \ket{i}$ is available: in other words, it is necessary to prepare a state in amplitude encoding beforehand. As a simple case, though, qRAMs allow for the preparation of a (sort of) equally weighted encoding $N^{-1/2} \sum_i \ket{i} \ket{x_i}$ in constant time, since the preparation of the equal superposition can be done in depth $1$.

The qRAM encoding can be exploited for quantum arithmetic as most of the encodings grounded on the basis representation. On the contrary, qRAMs are not useful for the preparation of amplitude encodings themselves. The next converter provides a way to prepare amplitude encodings.

\paragraph{Equally-weighted to amplitude.} Ref.~\cite{mitarai_quantum_2019} shows that it is possible to convert an equally-weighted encoding into an amplitude encoding, but the protocol requires measurements, it is successful under a given probability that is in general below $1$, and it needs $U_D$ to be unitary, where $U_D$ is the loading circuit producing the equally-weighted state. More precisely, let $[ d_i ]_{i=1}^N$ be a collection of binary fractions of maximal exponent $m$, namely $d_i = \sum_{k=1}^m d_i^{(k)} 2^{-k}$. Then there is a quantum algorithm that translates $N^{-1/2} \sum_i \ket{i} \ket{d_i^{(1)} \cdots d_i^{(m)}}$ into $\sum_i d_i \ket{i} \ket{0}$, up to a normalization factor. The algorithm requires $O(\poly m)$ single- and two-qubit gates, and one call to the inverse loading unitary $U_D^\dag$. It is successful with probability $N^{-1} \sum_i d_i^2$. Notice that the probability is asymptotically vanishing in $N$, for most cases of practical relevance, e.g. when they are sampled from a random variable $d$ such that $\mathbb E [d] \neq 1$ (and therefore $<1$).

\paragraph{Amplitude to equally-weighted.} Conversely, given a loading unitary $U_A$ in amplitude encoding, it is possible to produce the equally-weighted state that approximates it on $m$ binary digits, by resorting to $O(2^m)$ controlled applications of $U_A$ and to $O(2^m \log^2 N)$ single- and two-qubit gates with output fidelity $1- O(\poly 2^{-m})$~\cite{mitarai_quantum_2019}. The protocol is grounded on the Quantum Phase Estimation algorithm~\cite{kitaev_quantum_1995,nielsen_quantum_2010}.

\begin{figure}[t]
    \centering\small
    \begin{tabular}{c}
    {\includegraphics[trim={5.8cm 20.3cm 11cm 6.1cm},clip]{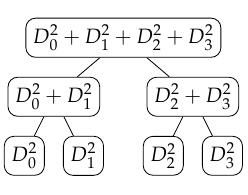}}
    \\(a)
    \end{tabular}
    \hfill
    \begin{tabular}{c}
    {\includegraphics[trim={6.5cm 15.5cm 3.3cm 11cm},clip]{Images/Binary_tree.pdf}}
    \\(b)
        \end{tabular}
    \caption{Two classical binary tree representations of a data set~\cite{araujo_configurable_2022}: (a) the state decomposition representation, and (b) the angle representation. The state decomposition can be built bottom up starting from a classical array, and it also applies to non-normalized data sets. The angle representation is specifically suited for quantum data loading, and can be derived by traveling the state decomposition tree top-down. Dashed nodes are redundant, since they can inferred from their sibling.}
    \label{fig:binary-tree}
\end{figure}

\paragraph{Classical binary tree structures.} Finally, in the attempt of accelerating the loading of data in the amplitude encoding, the advantage of preparing classical auxiliary data structures was explored. Such preparation has a classical overhead, that in principle may undermine the quantum speedups, unless this preparation is natural in the way data are generated, or the same data are reused multiple times, justifying the overhead. The classical binary trees, exemplified in Fig.~\ref{fig:binary-tree}, are by far the most widespread tool in this landscape.

The preparation of a binary tree costs $O(N)$ time on a classical computer~\cite{araujo2021divide, araujo_configurable_2022}.
The same binary tree structure also underlies the data preparation techniques already cited for the D\&C and bidirectional encodings~\cite{araujo2021divide,araujo_configurable_2022}. Refer to Subsection~\ref{subsec:encoding-ds} for depths.
The classical binary tree was shown to speedup also the classical computation, mostly due to the work of Ewin Tang~\cite{tang_quantum-inspired_2019, tang_quantum_2021}: indeed, the binary tree provides sampling access to the probability distribution associated to the data set. In other words, whenever a binary tree structure is available, the quantum algorithms suffer in principle the competition against a wider set of applicable classical algorithms.

\subsection{Distribution-specific loading techniques}
Under the hypothesis that the data distribution is known a priori, it may be possible to derive a more efficient loading technique. For example, Ref.~\cite{milek_quantum_2020} provides a technique to efficiently load discretized copulas in the amplitude encoding. Copulas~\cite{schmidt_coping_2006, nelsen_introduction_1999, mcneil_quantitative_2005} are multivariate distributions with uniform marginal probabilities, of practical relevance in finance~\cite{low_canonical_2013, yew_low_enhancing_2016}, since they allow for the decomposition of any joint distribution into univariates through Sklar's theorem~\cite{sklar_fonctions_1959, durante_topological_2013}.

Ref.~\cite{rattew_efficient_2021} shows how to load a normal distribution with a non-unitary circuit in amplitude encoding, with logarithmic asymptotic complexity and expected success rate independent of the number of qubits.

\subsection{Approximate data loading}
Given the high cost of data loading, the following alternative is being developed: approximating the input with a new data set that leads to a more compact unitary. Said approximations typically require a classical or hybrid preprocessing, that may be very computationally intensive, so that this option provides speedup only when the same data set is reused multiple time, thus absorbing the overhead of preprocessing. Besides this, on noisy devices, efficient data loading provide a second advantage, as they reduce the depth of circuits. Shallower circuits imply that higher portion of the limited quantum coherence time can be devoted to core computation rather than loading: in this setting, the interest does not lie in the asymptotic performance or overall time, but rather in the compression of the individually executed quantum circuits.

\paragraph{General unitary approximation techniques.} A first strategy is to build a (unitary) circuit that loads the data set exactly, and then approximate it. Indeed, good unitary approximation techniques are known, that reduce significantly the circuit depth. Approximate Quantum Compiling~\cite{madden_best_2022} is designed to find the best approximation for a given CNOT count, where the error is measured in terms of the Frobenius norm of the difference matrix. As other methods, unitary approximations become impractical above 7-8 qubits~\cite{madden_sketching_2022, ghosh_energy_2024}, due to barren plateaus, namely wide regions in the parameter space where the objective function is nearly flat, making optimizers fail in the identification of the minima. The piecewise AQC~\cite{ghosh_energy_2024} is a generalization designed for wider circuits.

\paragraph{Quantum machine learning.} Other important methods for approximate data loading originate in Quantum Machine Learning. Specifically, Quantum Generative Adversarial Networks (qGANs) play a key role. They were first applied for data loading in the domain of finance through the seminal work of Zoufal \textit{et al.}~\cite{zoufal_quantum_2019}. Since qGANs can load data in the amplitude encoding, their potential applications are countless~\cite{agliardi_quantum_2022}. The performance of qGANs, as well as some scaling limitations, are explored in our work~\cite{agliardi_optimal_2022,agliardi2022optimized}. A recent result~\cite{letcher_tight_2023} shows that appropriately designed qGANs are not affected by barren plateaus, which are otherwise a common issue in quantum machine learning.

\section{Data extraction}\label{sec:extraction}
By the term \textit{data extraction} we refer to the operation of retrieving the output of a quantum algoritm in terms of classical information.
Data extraction is related to the way the output is encoded by a quantum state. We demonstrate how Quantum Amplitude Estimation can be seen as an advanced data extraction technique.

The retrieval of classical information from a quantum state is performed through projective measurements. A projective measurement is described by an \textit{observable} $M$, namely a Hermitian operator on the state space~\cite{nielsen_quantum_2010}. Its spectral decomposition is $M = \sum_m P_m$, where $P_m$ is the projector onto the eigenspace of $M$ with eigenvalue $m$. After measuring a state $\ket{\psi}$, the probability to get the outcome $m$ is $p(m) = \mel**{\psi}{P_m}{\psi}$, and the quantum state gets modified by the measurement operation into $p(m)^{-\frac{1}{2}} \, P_m \ket{\psi}$ when the outcome $m$ occurred.

In practice, commercial quantum hardware typically offer the possibility of applying single-qubit measurements on a specific basis, such as the $X$ basis. The state can be appropriately rotated before measurements to project on other bases. A qubit measurement provides outcome $m=\pm 1$ according to the amplitude of the qubit in the state $\ket{\psi}$: if $\ket{\psi} = a_0 \ket{\psi_0} \ket{0} + a_1 \ket{\psi_1} \ket{1}$, the probability to get an output $-1$ is $\abs{a_0}^2$ and the probability of $+1$ is $\abs{a_1}^2$. In the rest of the paper, consistently with the literature in quantum computation, we have conventionally labelled the measurement outcomes $\{-1, +1\}$ as $\{0, 1\}$ respectively.

As a consequence, measurements access only part of the information encoded by a state, and partially destroy the state, coherently with Heisenberg's uncertainty principle. Consequently the state needs to be prepared multiple times and measured multiple times, to gain statistical significance over the possible measurement outcomes. This means that the whole circuit execution is repeated, giving rise to a number of \textit{shots}. The number of shots necessary to achieve a given accuracy $\epsilon$ in the outcome is called \textit{sampling complexity}, and together with the \text{circuit width} contributes to the estimation of the runtime of a quantum algorithm.

The complete characterization of the state of a quantum system is an operation known as \textit{quantum state tomography}, and requires in general a number of measurements exponential in the number of qubits~\cite{van_apeldoorn_quantum_2022}. As a consequence, quantum algorithm are designed in such a way that the relevant information are accessible through a limited number of projections, without need to fully characterize the state. This corresponds to obtaining the output in a specific encoding. In the remainder of the Section we discuss different extraction techniques according to the encoding itself.

\paragraph{From basis encoding.}
Suppose a quantum algorithm produces a state, such that the desired information is encoded in the basis of a subregister, namely in the form
\begin{equation}\label{eq:basis-output}
    \ket{\psi} = \ket{x} \ket{\phi},
\end{equation}
where $\ket{x}$ is a basis state representing the answer, while $\ket{\phi}$ is garbage. Then, a single measurement is sufficient to retrieve the desired output, in absence of noise. Some simple algorithms do effectively produce an output in the form of Eq.~\eqref{eq:basis-output}: it is the case of the Deutsch-Jozsa algorithm~\cite{deutsch_rapid_1992}, the Bernstein-Vazirani algorithm~\cite{bernstein_quantum_1997}, Simon's algorithm~\cite{simon_power_1997}, Grover's search algorithm when the solution is unique~\cite{grover_fast_1996}.

\paragraph{From basis encoding with the highest probability.}
The previous case is very peculiar, and in algorithm design it is infrequent to be able to encode the output in the basis state form so that it can be measured with certainty. More often though, one can guarantee that the correct answer is encoded by a basis state which is the most likely to be measured, namely the mode. In formulas, before measurement the quantum algorithm produces a state in the form
\begin{equation} \label{eq:basis-highest}
\ket{\psi} = \sum_x a_x \ket{x} \ket{\phi_x},
\end{equation}
where $\ket{x}$ are basis states, $\ket{\phi_x}$ are garbage, and the correct answer is the most likely of the values $x$, namely an $\bar x$ such that $\abs{a_{\bar x}}^2 =: p$ is higher than any other $\abs{a_{x}}^2$.

\begin{remark}\label{rem:success-amplification}
By repeated experiments, the probability of failing in obtaining $\bar x$ decreases exponentially. More precisely, let $M_S$ be the \textit{mode} of the measurement output of the first register in Eq.~\eqref{eq:basis-highest}, after repeating the execution $S$ times. Then
$$ \mathbb P (M_S = \bar x) \geq 1-2k R^S, $$
where $k$ is the amount of $x$ values with non-zero probability and $R$ is a characteristic of the distribution, always dominated by $1$~\cite[Thm.~4]{dutta_mode_2010}. It should be noted here that other sources use the median instead of the mode (e.g.~\cite[Appendix~F]{rebentrost_quantum_2018-1}), leveraging on the so-called `median lemma'~\cite[Lemma~1]{nagaj_fast_2009}, but this requires the additional hypothesis that $p>\frac{1}{2}$ and typically leads to worse constants.
\end{remark}

\paragraph{From basis encoding with high probability.} In other algorithms, the pattern is that the output state before measurement is again like the one in Eq.~\eqref{eq:basis-highest}, where this time the correct answer is not the \textit{most} likely, but \textit{one of} the highly probable, even in absence of error. A classical post-processing must then determine the correct solution among the candidates. An example of algorithm that falls into this category is the period finding~\cite[§5.4.1]{nielsen_quantum_2010} used in Shor's factoring algorithm.

\paragraph{From generalized amplitude encoding -- naif approach.} Another class is that of algorithms whose answer is encoded by the amplitude of a given basis state or set of basis states. The typical case is that the algorithm, before measurement, produces a state
\begin{equation}\label{eq:est-from-ampl}
    \ket{\psi} = a_0 \ket{0} \ket{\phi_0} + a_1 \ket{1} \ket{\phi_1},
\end{equation}
where the number of interest is $\abs{a_1}^2$. Since the measurement of the first qubit outputs $1$ with probability $\abs{a_1}^2 =: p$, one can define a random variable $X$ to be the outcome of the measurement, namely a Bernoulli with parameter $p$. The mean $\bar X_S$ of $S$ independent and identically distributed random variable like $X$ is an estimator for $p$. By the central limit theorem, $\bar X_S$ has a variance of $p (1-p) S$, implying that the sampling complexity to guarantee an absolute error $\epsilon$ with confidence $\alpha$ is asymptotically $S=O(\epsilon^{-2})$. More precisely,
$$\mathbb P \left( \abs{\bar X_S - p} < \epsilon \right) \geq \alpha$$
is asymptotically equivalent to
$$S = \epsilon^{-2} \, p(1-p) \, \left[ \Phi^{-1} \left( \frac{1+\alpha}{2} \right) \right]^2 $$
when $\epsilon \to 0$, where $\Phi$ is the cumulative distribution function of a standard normal.

\paragraph{From generalized amplitude encoding -- Quantum Amplitude Estimation.}
The Quantum Amplitude Estimation (QAE) is a method that accelerates the estimation of the magnitude of the amplitude of a state \cite{maronese2024quantum}. Compared to the previous naif approach it achieves a quadratic speedup, under the hypothesis that the state $\ket{\psi}$ is produced by a unitary circuit $F$, and that such unitary is known.
Originally due to Brassard \textit{et al.}~\cite{brassard_quantum_1998, brassard_quantum_2002}, it is formalized in the following Proposition. The reader can also refer to Ref.~\cite{rebentrost_quantum_2018-1} for a proof.

\begin{figure}
    \centering\small
    \includegraphics{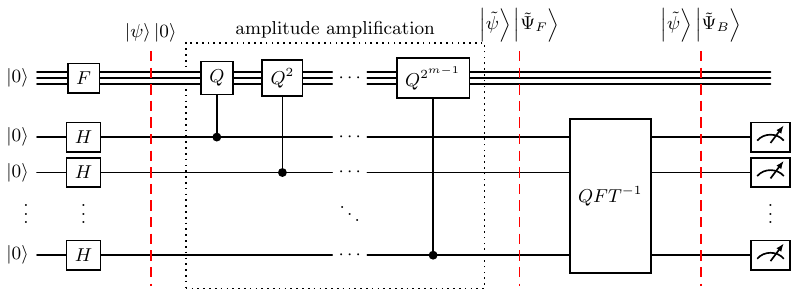}
    \caption{The Quantum Amplitude Estimation circuit. The circuit preparation requires a unitary operator $F$ preparing the state $\ket{\psi}$ in Eq.~\eqref{eq:est-from-ampl}, where the quantity of interest is $\mu = \abs{a_1}^2$. Call $x = \overline{.x_0 \dots x_{m-1}}$ the approximation of $\mu$ in $m$ binary digits. The operator $Q = Q(F)$ is obtained through a standard procedure from $F$. The amplitude amplification stage produces a state $\ket{\tilde \Psi_F}$ that is close to the Fourier encoding $\ket{\Psi_F}$ of $x$. Assuming the simplified case where $\ket{\tilde \Psi_F} = \ket{\Psi_F}$, the state $\ket{\Psi_F}$ can be then converted into a basis encoding $\ket{ \Psi_B} = \ket{x_0 \dots x_{m-1}}$ through the QFT as explained in Subsec.~\ref{subsec:load-point}. Finally, $\ket{\Psi_B}$ is measured to obtain the classical number $\overline{x_{m-1} \dots x_0}$, which is finally mapped into the estimator $\hat \mu$ for the desired quantity $\mu$ trough a simple trigonometric transformation. The whole algorithm is applicable also when $\ket{\tilde \Psi_F}$ and $\ket{\Psi_F}$ do not match exactly, and in this case the estimation is not obtained with certainty, but with high probability. Details on how to obtain $Q$ from $F$ and how to map $\overline{x_{m-1} \dots x_0}$ to $\hat \mu$, can be found in the references, e.g.~\cite{rebentrost_quantum_2018-1}. Notice that the amplification phase may alter the original state $\ket{\psi}$ into $\ket{\tilde \psi}$. The alteration scheme though is very peculiar, giving rise to a set of variants of QAE that preserve the input state~\cite{rall_amplitude_2022}.}
    \label{fig:qae}
\end{figure}

\begin{prop}[QAE]\label{prop:qae}
Suppose a unitary operator $F$ is given, such that $\ket{\psi} = F \ket{0}$, where $\ket{\psi}$ is the state of interest in Eq.~\eqref{eq:est-from-ampl}.
Then, the Quantum Amplitude Estimation method builds a circuit whose measurement, once post-processed, provides an estimator $\hat \mu$ for $\mu := \abs{a_1}^2$ with the highest probability:
$$\mathbb P \left( \abs{\hat\mu - \mu} \leq 2^{-m} \right) \geq \frac{8}{\pi^2}.$$

The algorithm employs $m$ qubits in addition to those necessary for $\ket{\psi}$, and one shot requires $O(2^m)$ applications of $F$.

Additionally, the algorithm can be refined to any success probability $\alpha > \frac{8}{\pi^2}$, namely
$$\mathbb P \left( \abs{\hat\mu - \mu} \leq 2^{-m} \right) \geq \alpha,$$
with $O \left(2^{-m} \log (1-\alpha)^{-1} \right)$ queries to $F$.
\end{prop}

In our vocabulary, the Proposition states that QAE translates the problem of extraction of outputs from amplitude encoding, into extraction from basis encoding with the highest probability (and hence with arbitrary probability, by resorting to the techniques introduced before). More in detail, Fig.~\ref{fig:qae} shows that QAE assumes the availability of an amplitude encoding unitary $F$ and obtains an intermediate QFT encoding, which is then converted into a basis encoding through QFT.

The query complexity of QAE scales as $O(\epsilon^{-1})$, where $\epsilon$ is the accepted error threshold ($\epsilon = 2^{-m}$ in the previous Proposition). Compared to the naif version of the extraction from amplitude encoding, that scales as $O(\epsilon^{-2})$, QAE achieves the claimed quadratic speedup.

In the attempt to reduce the circuit depth and width of QAE, multiple variants were proposed, including the Iterative QAE~\cite{grinko_iterative_2021}, the Chebyshev QAE~\cite{rall_amplitude_2022}, and the Dynamic QAE~\cite{ghosh_energy_2024}. Ref.~\cite{ghosh_energy_2024} also provides a summary comparison of a wider range of variants. We also report that Ref.~\cite{rall_amplitude_2022} achieved the remarkable result of formalizing multiple QAE methods under a common mathematical framework.

\section{Application: Quantum-based Monte Carlo}\label{sec:montecarlo}
\begin{figure}[t]
\small \centering
\includegraphics{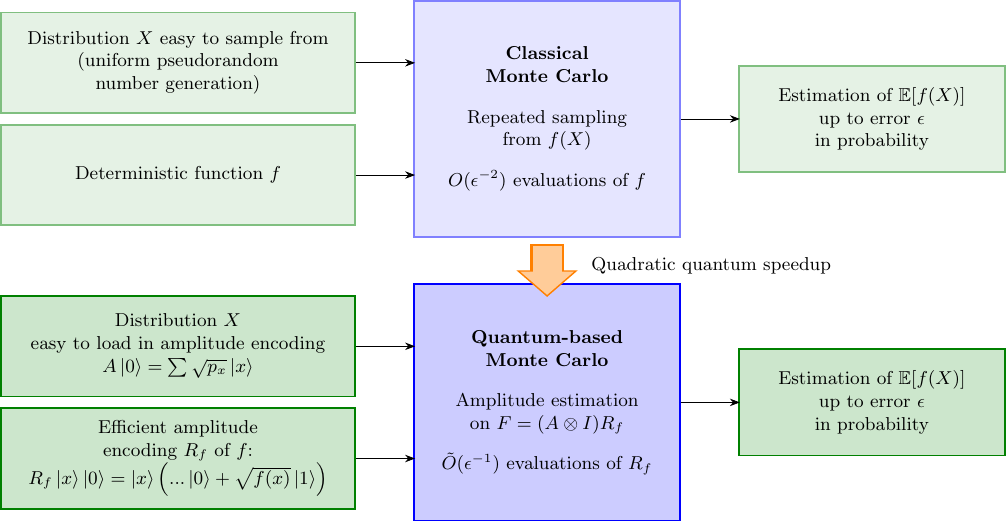}
\caption{A schematic comparison between classical and quantum-based Monte Carlo techniques, showing the inputs and outputs of the methods.}\label{fig:montecarlo}
\end{figure}

Quantum Amplitude Estimation techniques were introduced in the previous Section as a tool for data extraction from amplitude encoding. On the other hand, QAE can be exploited as a quantum analog of Monte Carlo methods, thanks to Montanaro's work~\cite{montanaro_quantum_2015}. Let us clarify here this relationship, also depicted in Figure~\ref{fig:montecarlo}.

A Monte Carlo method can be described as a couple $(X, f)$ where (a) $X$ is a random variable valued in a space $D$, and $X$ can be easily\footnote{Practically speaking, the easy sampling translates into the availability of a sufficiently long sequence of pseudo-random numbers that behave in agreement with the desired distribution law. Most commonly, $X$ is a uniform variable valued in a subset $D$ of $\mathbb R^n$.} sampled from, and (b) $f: D \to \mathbb R$ is a deterministic function, such that $Y=f(X)$ is an unknown distribution. The objective of the Monte Carlo method in this simplified version, is to estimate $\mathbb E[f(X)]$, repeating the function evaluation on $S$ samples. By the central limit theorem, the number of samples (and of queries) needed to obtain a precision $\epsilon$ in probability, scales as $O(\epsilon^{-2})$. For example, if $X$ is uniform on a domain $D$, $f$ is integrable on $D$ (and 0 elsewhere), then the Monte Carlo method provides an estimator for the integral $\frac{1}{\abs{D}} \int_D f$.

\begin{figure}
    \centering\small
    \includegraphics{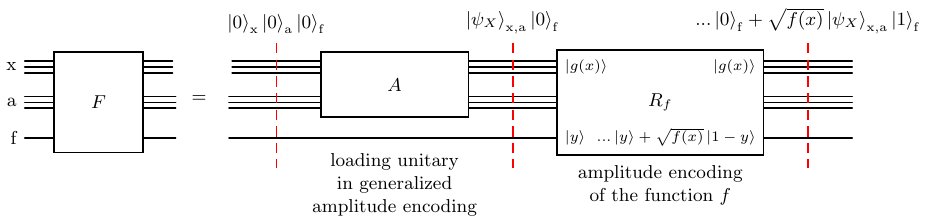}
    \caption{The operator $F$ in quantum-based Monte Carlo simulations is the combination of two operators $A$ and $R_f$, where the former loads the distribution $X$ into the first register $\ket{\cdot}_{\mathrm{x}}$ in the amplitude encoding (possibly in a generalized sense, with an entangled auxiliary register $\ket{\cdot}_{\mathrm{a}}$, as defined in Subsec.~\ref{subsec:encoding-ds} and described in Eq.~\ref{eq:montecarlo-x}), while the latter is an amplitude encoding of $f$ (Subsec.~\ref{subsec:encoding-fn}).}
    \label{fig:montecarlo-f}
\end{figure}

Now, let us move to the quantum algorithm method. Assume that $X$ is discrete, say $[ x_i ]_i$ are the $2^m$ mass points and $[ p_i ]_i$ the respective probabilities. Consider a generalized amplitude encoding of $X$ in the following sense:
\begin{equation}\label{eq:montecarlo-x}
\ket{\psi_X}_{\mathrm{x,a}} = \sum_{i=0}^{2^m-1} \sqrt{p_i} \ket{x_i}_{\mathrm{x}} \ket{\phi_i}_{\mathrm{a}} = \sum_{i=0}^{2^m-1} \sqrt{p_i} \ket{\psi_i}_{\mathrm{x,a}}
\end{equation}
for some array $[ \ket{\phi_i} ]_i$ of states belonging to an auxiliary register. Also, suppose there exists a unitary $A$ that loads $\ket{\psi_X}$, namely $\ket{\psi_X} = A \ket0$. Finally, assume that an operator $R_f$ is the amplitude encoding of $f$, i.e.:
\begin{equation}\label{eq:rf}
R_f \ket{x}_{\mathrm{x}} \ket{0}_{\mathrm{f}} = \ket{x}_{\mathrm{x}} \left(\sqrt{1-f(x)} \ket{0}_{\mathrm{f}} + \sqrt{f(x)} \ket{1}_{\mathrm{f}} \right).
\end{equation}
Now, Eq.~\eqref{eq:rf} implies that $F := (A \otimes I) R_f$ produces the state
$$F \ket{0}_{\mathrm{x}} \ket{0}_{\mathrm{a}} \ket{0}_{\mathrm{f}} = \sum_{i=0}^{2^m-1} \sqrt{p_i \, f(x_i)} \ket{\psi_i}_{\mathrm{x,a}} \ket{1}_{\mathrm{f}} + (...) \ket{0}_{\mathrm{f}}$$
which in turn allows for the application of QAE according to Prop.~\ref{prop:qae}, thus providing an estimate of
$$\sum_{i=0}^{2^m-1} p_i \, f(x_i) = \mathbb E [f(X)]$$
in $O(\epsilon^{-1})$ queries.
The reduction of the \textit{query complexity} from $O(\epsilon^{-2})$ of classical Monte Carlo methods to $O(\epsilon^{-1})$ of QAE is the celebrated quadratic quantum speedup on Monte Carlo simulations. To inherit said speedup in terms of runtime, some more assumptions are needed, as we comment in the next remarks.

\begin{remark}[Time comparison against direct summation]
In the quantum case, the variable $X$ is assumed to be discrete (as much as, at a closer look, the pseudo-random sequence classically generated \textit{is} discrete). To justify the application of quantum-based Monte Carlo approximation, though, the number $2^m$ of mass values $x_i$ in the distribution must be bigger than the number of query evaluations required by the QAE: otherwise, the direct exact calculation of
$\mathbb E [f(X)] = \sum_i p_i \, f(x_i)$ on a classical computer would be less costly, as it requires $\Theta(2^m)$ time. In other words, $\epsilon^{-1} =O(2^m)$ must hold, otherwise the quantum method is certainly not convenient. The last equality takes proper meaning when $X$ is the discretization of a target continuous random variable, so that $m$ can be varied indeed.
\end{remark}

\begin{remark}[Time comparison against classical Monte Carlo]
To achieve a speedup not only in terms of query complexity but also in time complexity, the circuit depth of $F$ must keep controlled when $m$ increases. If $c_F(m)$ is the depth complexity of $F$, then the time scaling of the quantum-based Monte Carlo is $O(c_F(m) \, \epsilon^{-1})$, that compares to the classical Monte Carlo scaling of $O(c_f(m) \, \epsilon^{-2})$ where $c_f(m)$ is the evaluation time\footnote{In a modern approach to classical computational theory, the evaluation time $c_f$ would be generally considered constant, as classical arithmetic is extremely efficient on present computer architectures. Nevertheless, properly speaking, $c_f$ is a function of the number of significant digits required, so that it is not constant, but should explicitly dependent on $\epsilon$ rather than $m$. Such refinements in the present theory are left for future work. Here we limit to keep $c_f$ manifest, so that its contribution can be matched with that of $c_F$ in the quantum case.} of $f$ on a single data point out of $2^m$. A time speedup against the classical Monte Carlo exists only if $\frac{c_F(m)}{c_f(m)} \ll \epsilon^{-1}$, and the quadratic advantage requires $\frac{c_F(m)}{c_f(m)} =O(1)$.
\end{remark}

\begin{remark}[Time comparison summary]
Combining the two previous remarks, the conditions $m \gg - \lg \epsilon$ and $c_F(m) \ll 2^m$ are reasonable requirements for a quantum-based Monte Carlo method. The former determines how big the sampling distribution should be in comparison with the desired error threshold, and gives bounds on the quantum hardware scale that should be employed for the practical application of the methods described herein. The latter instead is a condition on the ability to design an effective quantum circuit for the problem. In practice $c_F(m) \ll 2^m$ typically translates into analog conditions on the time $c_A(m)$ of the loading unitary $A$, and $c_R(m)$ of the processing unitary $R_f$. Now, the fact that $c_A$ scales well with the number of mass points $2^m$, corresponds to the hypothesis that $X$ is an easy distribution to sample from, imposed in classical Monte Carlo: as an extreme example, uniform distributions, that are typically the target of pseudo-random generation in classical computing, are also trivially loaded into a quantum computer through $H^{\otimes m}$. The condition of $c_R$ scaling well with $m$, on the contrary, is more subtle, and does not find a perfect match with the classical case, where the evaluation of a function is not affected by the size of the potential domain, but only by the desired accuracy.
\end{remark}

As a final comment, let us say that the technique of quantum-based Monte Carlo simulations naturally extends to encodings with mappings in the sense of Subsec.~\ref{subsec:encoding-qv-anticipation} and Fig.~\ref{fig:g-encoding}, as examined in the next Remark.
\begin{remark}\label{rem:montecarlo-with-mapping}
Let $X$ be a random variable taking values in a discrete domain $\mathcal{D}=\{x_i\}_{i=0}^{2^n-1}$. Let $g$ be a bijection from $\mathcal{D}$ to $\{0, ..., 2^{n-1} \}$. Then, the distribution $X$ can be $g$-encoded in the amplitudes in the generalized form
$$\ket{\psi_X}_{\mathrm{a,x}} = \sum_{i} p_i \ket{g(x_i)}_{\mathrm{x}} \ket{\psi_i}_{\mathrm{a}}$$
where $p_i = \mathbb P (X=x_i)$. The quantum-based Monte Carlo technique described above keeps working under this generalization, as long as $R_f$ is compatible with $g$ in the following sense:
\begin{equation*}
R_f \ket{g(x)}_{\mathrm{x}} \ket{0}_{\mathrm{f}} = \ket{g(x)}_{\mathrm{x}} \left(\sqrt{1-f(x)} \ket{0}_{\mathrm{f}} + \sqrt{f(x)} \ket{1}_{\mathrm{f}} \right),
\end{equation*}
for all $x \in \mathcal D$.
Said condition is satisfied for instance by some floating-point encodings~\cite{doriguello_quantum_2022,agliardi_floating_2023}.
\end{remark}

\section{Conclusions}\label{sec:conclusions}
We formalized the concept of quantum data encoding and explored different techniques of data loading, encoding conversion and data extraction. In this framework, we described the QFT as a data encoding conversion routine. Similarly, we depicted QAE as a data extraction routine, distancing ourselves from the most common description of QAE as a quantum-based Monte Carlo integration tool. We have shown how encodings can also be used to describe intermediate stages in the circuit of QAE. Finally, as an application, we reinterpreted QAE as a Monte Carlo method, and showed that in such translation quantum encodings are again helpful for a high level understanding of the circuit composition.

Data encodings are the interfaces connecting multiple components of a quantum circuit, and as such, with this work, we contributed to a conceptual representation of circuits as modular blocks.

The present collection of data encodings and associated loading, conversion, and extraction techniques, is far from being thorough. The conceptual framework here proposed therefore needs enrichment and validation on a wider spectrum of building blocks as well as composite workloads. Nonetheless, we believe that the new perspective may result useful for the understanding, description and design of complex quantum circuits, as well as in the definition of high-level quantum programming languages.

\section*{Acknowledgments}
This work is the result of a gradual conceptualization process triggered by many conversations. G.A. is thankful to the participants of the quantum computing course organized by CRUI (Conferenza dei Rettori delle Università Italiane) in 2021, as well as to Omar Shehab, Corey O'Meara, Kumar Ghosh, Kavitha Yogaraj, and Andrea Delgado for inspiring discussions. E.P. gratefully thanks the participants of the Study Day on Open Problems in Quantum Machine Learning  held in Milan in October 2022 for having openly shared their thoughts. E.P. acknowledges the project Qxtreme for having partially supported this research.

\section*{Conflicts of interests}
The authors declare no conflict of interest.

\bibliography{biblio}{}
\bibliographystyle{ieeetr}

\end{document}